\documentclass[onecolumn]{aastex63}
\usepackage[encapsulated]{CJK}
\usepackage{mathrsfs}
\usepackage{subfigure}
\usepackage{epstopdf}
\usepackage{amsmath}
\usepackage{longtable}
\usepackage{booktabs}
\usepackage{hyperref} 
\usepackage{overpic}
\usepackage{threeparttable}
\usepackage{subfigure}
\usepackage{diagbox}
\usepackage{rotating}

\usepackage{lineno}

\newbox\grsign \setbox\grsign=\hbox{$>$} \newdimen\grdimen \grdimen=\ht\grsign
\newbox\laxbox \newbox\gaxbox
\setbox\gaxbox=\hbox{\raise.5ex\hbox{$>$}\llap
     {\lower.5ex\hbox{$\sim$}}}\ht1=\grdimen\dp1=0pt
\setbox\laxbox=\hbox{\raise.5ex\hbox{$<$}\llap
     {\lower.5ex\hbox{$\sim$}}}\ht2=\grdimen\dp2=0pt
\shorttitle{Spiral structure of the Milky Way}
\shortauthors{Xu et al.}

\definecolor{malachite}{rgb}{0.34, 0.7, 0.22}

\begin{document}

\title{What does the Milky Way look like?}
   
\correspondingauthor{Ye Xu}
\email{xuye@pmo.ac.cn}
   
\author{Y. Xu} 
\affiliation{Purple Mountain Observatory, Chinese Academy of Sciences, Nanjing 210023, People's Republic of China}
\affiliation{School of Astronomy and Space Science, University of Science and Technology of China, Hefei 230026, People's Republic of China}

\author{C. J. Hao}
\affiliation{Purple Mountain Observatory, Chinese Academy of Sciences, Nanjing 210023, People's Republic of China}
\affiliation{School of Astronomy and Space Science, University of Science and Technology of China, Hefei 230026, People's Republic of China}

\author{D. J. Liu}
 \affiliation{Purple Mountain Observatory, Chinese Academy of Sciences, Nanjing 210023, People's Republic of China}
\affiliation{School of Astronomy and Space Science, University of Science and Technology of China, Hefei 230026, People's Republic of China}
\author{Z. H. Lin}
 \affiliation{Purple Mountain Observatory, Chinese Academy of Sciences, Nanjing 210023, People's Republic of China}
\affiliation{School of Astronomy and Space Science, University of Science and Technology of China, Hefei 230026, People's Republic of China}
\author{S. B. Bian}
\affiliation{Purple Mountain Observatory, Chinese Academy of Sciences, Nanjing 210023, People's Republic of China}
\affiliation{School of Astronomy and Space Science, University of Science and Technology of China, Hefei 230026, People's Republic of China}
\author{L. G. Hou}
\affiliation{National Astronomical Observatories, Chinese Academy of Sciences, 20A Datun Road, Chaoyang District, Beijing 100101, People's Republic of China}
\author{J. J. Li}
\affiliation{Purple Mountain Observatory, Chinese Academy of Sciences, Nanjing 210023, People's Republic of China}
\affiliation{School of Astronomy and Space Science, University of Science and Technology of China, Hefei 230026, People's Republic of China}
\author{Y. J. Li}
\affiliation{Purple Mountain Observatory, Chinese Academy of Sciences, Nanjing 210023, People's Republic of China}

 
\begin{abstract}
In spite of much work, the overall spiral structure morphology of the Milky Way 
remains somewhat uncertain. In the last two decades, accurate distance 
measurements have provided us with an opportunity to solve this issue. Using 
the precise locations of very young objects, for the first time, we propose 
that our Galaxy has a multiple-arm morphology that consists of two-arm 
symmetry (the Perseus and Norma Arms) in the inner parts and that extends 
to the outer parts, where there are several long, irregular arms (the Centaurus, 
Sagittarius, Carina, Outer, and Local Arms).
\end{abstract}

\keywords{Milky Way Galaxy -- Galactic structure -- Trigonometric parallax}

%

\section{Introduction}
\label{intro}

Determining the detailed spiral structure of the Milky Way (MW) has long
been a difficult issue in astronomy. Since we are deeply embedded in
the Galactic disk, there are always multiple structural features
superimposed along the observational line of sight.
For a spiral galaxy, there may be two different components of spiral
arms \citep[][and references therein]{dobbs2014}. One is composed of a
spiral pattern indicated by the distribution of the older stellar
population, and the other is a spiral picture traced by diffuse or dense 
interstellar gas and young objects, e.g., high-mass star-forming regions 
(HMSFRs), massive O--B stars, H\textsc{ii} regions, young open 
clusters (YOCs), etc.

Pioneering work was made by Morgan and his colleagues in the 1950s. 
They found three short spiral-arm segments in the solar neighborhood 
using spectroscopic parallaxes of high-mass stars \citep{morgan1952,morgan1953}. 
Soon after, larger-scale spiral structure, extending almost across the 
entire Galactic disk, was mapped with H\textsc{I} survey data using 
kinematic methods \citep[e.g.,][]{van1954,kerr1957,bok1959,burton1970}. 
Later, however, the distances of H\textsc{I} clouds, derived from 
kinematic methods, were found to have large uncertainties due to 
noncircular (peculiar) motions. Thus, the H\textsc{I} results in these 
early studies are not very reliable.
Photometric methods are much more accurate than kinematic methods, 
although they can only be used to determine objects at distances up to 
$\sim$ 2 kpc, i.e., much smaller than the size of the MW. For this reason, 
kinematic methods are still widely used to study the entire MW.
By analyzing a number of H\textsc{ii} regions with photometric and/or 
improved kinematic distances, a paradigmatic map of the Galaxy's spiral 
arms was made by~\citet{georgelin1976}. They first proposed that the MW 
has four major arms.
This picture was then frequently updated by using other spiral
tracers, such as molecular clouds~\citep[e.g.,][]{burton1978,dame1987,
dame2001,sun2015,du2016}, star-forming
complexes~\citep[][]{russeil2003}, a larger sample of H\textsc{ii}
regions~\citep[e.g.,][]{paladini2004,hou2014}, H\textsc{I}  gas~\citep[e.g.,][]{levine2006}, etc.,
most of which relied on kinematic methods. 

Although kinematic methods are being improved all the time, they 
sometimes cause significant uncertainties in the locations of sources. 
Therefore, the debate continues over such basic facts as the existence 
of some spiral arms, the number of arms, and the size of the MW. 
Measuring distances as accurately as possible using spiral tracers is key 
to settling these disputes and correctly uncovering the Galaxy's spiral structure.
Recently, substantial progress in tracing the Galactic structure using 
young objects has been achieved. On the one hand, Very Long Baseline 
Interferometry (VLBI) can yield trigonometric parallax accuracies down to 
a few $\mu$as \citep[e.g.,][]{xu2006,hachisuka2006,sanna2017}, allowing 
precise distance measurements toward masers associated with HMSFRs
throughout the Galaxy. Focusing on mapping the MW, the Bar and Spiral 
Structure Legacy (BeSSeL) Survey~\citep{brunthaler2011} and the VLBI 
Exploration of Radio Astrometry (VERA) array~\citep{vera2020} have 
measured accurate distances for up to $\sim$200 masers.
These measurements indicate that the MW is a four-arm spiral, with 
some extra arm segments and spurs \citep[][hereafter R19]{reid2019}.
On the other hand, a large amount of young stars provided by the 
\emph{Gaia} mission has densified the spiral-arm segments traced by 
masers, which also help determine the Galactic spiral structure in the solar
neighborhood~\citep{xu2018a,xu2018b,xu2021a,hao2021,hou2021,poggio2021}.

External spiral galaxies can also act as a mirror to help us understand 
Galactic morphology better as the MW is one of trillions of galaxies 
in the observable universe. Pictures of external spiral galaxies show 
that there are largely three distinct types of morphology. In two extreme 
cases, grand-design spiral galaxies are highly symmetric, characterized 
by clear, long, and symmetric spiral arms, whereas flocculent ones are 
fragmented, consisting of many short, irregular, and patchy segments. An
intermediate type lies in between, so-called multiple-arm spiral
galaxies whose main characteristic is an inner two-arm symmetry and
several irregular arms in the outer parts~\citep{elmegreen2014}.
Different from flocculent galaxies, both grand-design and multiple-arm 
galaxies have two prominent, symmetric arms in their inner regions, and 
almost no external galaxies present four spirals extending from their 
centers to their outer regions~\citep{elmegreen1982,elmegreen1987,elmegreen1995}. 
Today, it is widely accepted that the MW galaxy has four continuous 
spiral arms extending outward from the inner Galaxy to distant outer 
regions~\citep[e.g.,][]{georgelin1976,reid2019,minniti2021}. If that is the 
case, the MW may be an atypical galaxy in the universe.

In the past few years, the number of young objects in the MW with
precisely measured distances has increased significantly. It would be
constructive to combine these high-quality data together to provide a 
better understanding the MW's spiral structure.
In this contribution, largely following the previously developed
spiral-arm-fitting approach presented by R19, we have synthesized the
available data set of spiral tracers with precisely measured distances,
including the parallax measurements of HMSFR masers from VLBI
observations, massive O--B2-type stars, and YOCs from the \emph{Gaia} 
mission, aiming to uncover the real image of our archetypal Galaxy.

\section{Data}
\label{data}
Currently, VLBI maser parallax measurements are mainly carried out in the
Northern Hemisphere, with uncertainties of typically $\sim$20$~\mu$as, 
while a number of observations have uncertainties down to $\sim$10$~\mu$as 
or better, allowing reliable distances to be determined for objects located at 
the Galactic Center (GC) and beyond~\citep{zhang2013,sanna2017,reid2019,xu2021b}.
On the other hand, \emph{Gaia} Data Release 3 (DR3) data have 
uncertainties of order 20--30~$\mu$as, which enable us to reveal the spiral 
structure within $\sim$5 kpc of the Sun.
In addition to using these distances to trace the spiral structure directly, 
ascertaining arm tangencies are also a good way to determine the locations 
of spiral arms, which are identified from the distributions of stars, interstellar 
gas, dust, and star-formation sites in the Galactic plane~\citep{hou2015}.

\subsection{Masers}
VLBI at radio wavelengths has been used to detect a large amount of masers 
associated with HMSFRs, and they can be used to map the spiral structure of 
the MW from the solar neighborhood to the GC and beyond.
Table~\ref{tab:masers} lists the coordinates and parallaxes of 204 HMSFR
masers measured with VLBI techniques, including the National Radio Astronomy 
Observatory's Very Long Baseline Array, the Japanese VLBI Exploration of Radio 
Astrometry project, the European VLBI Network, and the Australian Long Baseline 
Array. For the masers listed in Table~\ref{tab:masers}, 199 were summarized by 
R19, and five have been newly reported by~\cite{xu2021b} and \cite{bian2022} 
in the past two years. The typical parallax accuracy of the masers is 
$\sim$20~$\mu$as, and 19 of them even have accuracies of 10~$\mu$as or better.

\subsection{O--B2-type stars}
High-mass stars are generally located near their birthplaces, and thus they can 
also be used to trace the Galactic spiral structure. Employing the same method 
as used by \cite{gaia2022}, we adopted the effective temperature, spectral type, 
distance from the galactic plane, and the astrometric fidelity indicator, 
$f_{a}$~\citep{rybizki2022}, to select O--B-type star candidates from 
\emph{Gaia} DR3~\citep{gaia2022}. We only use stars with a parallax precision 
better than 20\%, i.e., \texttt{parallax\_over\_error} $>$ 5. After applying this 
criterion, we obtained a total of 495\,965 O--B stars.

Then, RR-Lyrae stars identified by \texttt{vari\_rryrae} in \emph{Gaia} DR3 were 
removed. Next, we filter the tangential velocity $\nu_{\rm tan} = A_{\nu}(\mu_{\alpha *}^{2} 
+ \mu_{\delta}^{2})^{1 / 2} / \varpi$ with  $\nu_{\rm tan} < 180$ km s$^{-1}$, the 
same as~\cite{gaia2018}, where $\mu_{\alpha *}$ and $\mu_{\delta}$ are the 
proper motions in the R.A. and decl., respectively, $\varpi$ is the parallax, and 
$A_{\nu}$ = 4.74~ km~yr~s$^{-1}$.  After filtering based on the above criteria, 
488\,449 O--B-type stars are left. In this work, we only consider stars hotter 
than 20\,000~K, which corresponds to O--B2-type stars whose typical ages 
are younger than 20~Myr \citep[][]{chen2013}. The final sample is composed 
of 23\,807 O--B2 stars. We perform a parallax zero-point correction for all 
O--B2-type stars in the sample, as indicated by~\citep{lindegren2021}. 
The median standard error of the parallaxes of the O--B2-type stars is 17~$\mu$as.

\subsection{Open clusters}
Since YOCs ($\textless$ 20 Myr) can trace the spiral arms well~\citep{hao2021}, 
we only use OCs younger than 20 Myr as tracers in this work. In recent years, 
taking full advantage of the high-precision astrometric data provided by \emph{Gaia}, 
numerous studies have identified thousands of OCs in the MW, and their ages cover 
a wide range, from several megayears to a few gigyears. Based on many previous works, 
\cite{hao2021} compiled a catalog containing more than 3\,700 OCs using 
\emph{Gaia} Early Data Release 3. Soon after, 704 and 628 newly found OCs in 
the \emph{Gaia} area were reported by \cite{hao2022} and \cite{castro2022}, 
respectively. In this work, we have synthesized the above three catalogs and 
obtained a large sample of 5\,021 OCs, of which 1\,011 OCs have ages younger 
than 20 Myr. We apply a parallax zero-point correction for the member stars 
of each OC, and then use the average parallax of all members as the OC's 
parallax~\citep{lindegren2021}. Similar to the O--B2-type star sample, we 
only use YOCs with parallax accuracies better than 20\%, resulting in a final 
number of 981 YOCs. The median standard error of parallax of YOCs is 
$\sim$23 $\mu$as.

\section{Objects Assigned to Spiral Arms}
\label{methods}
\subsection{Masers}
The spiral arms of the MW show quasi-continuous structure in CO and
H{\sc I} in Galactic longitude--local standard of rest (LSR) velocity 
($l$--$v$) maps. We initially assign masers to spiral arms by comparing 
the $(l, v)$ positions of the sources with the $(l, v)$ loci of arm segments 
provided by~\cite{reid2016}. Although a plan view of the MW from 
$l$--$v$ plots cannot accurately place HMSFR masers into spiral arms, 
one can, in most cases, assign HMSFR masers to spiral arms by their 
association with CO and H{\sc I} emission features. In addition, the 
Galactic latitudes and parallax distances can be helpful for confirming 
the arm assignments of HMSFR masers and avoiding any anomalous 
kinematic effects.

\begin{figure*}[htbp]
	\center
	\includegraphics[width=0.8\textwidth]{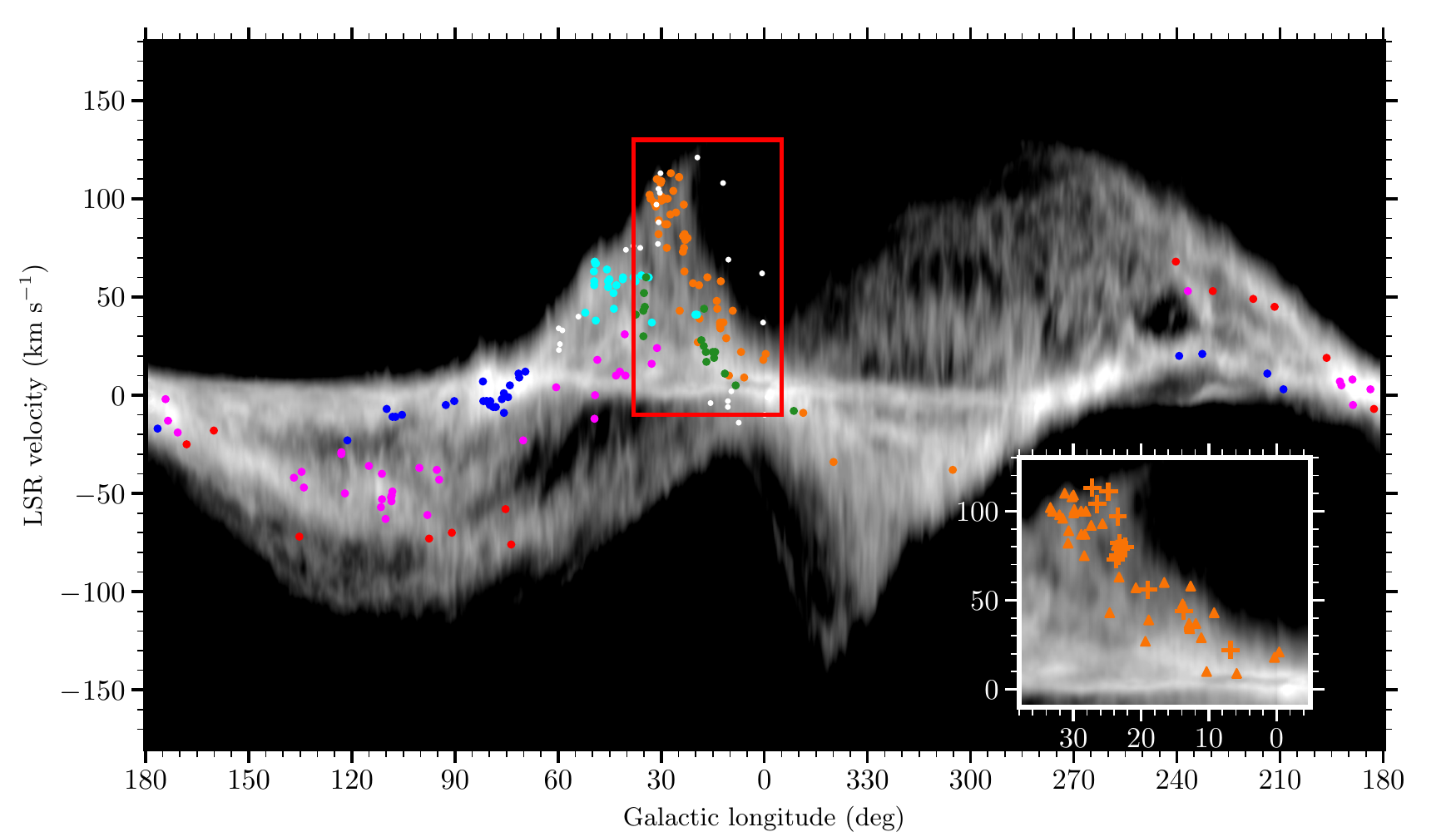}
	\caption{H{\sc I} emission as a function of LSR radial velocity and Galactic longitude. The colored dots are HMSFR masers with parallax measurements: Norma Arm (orange), Sagittarius Arm (cyan), Carina Arm (green), Local Arm (blue), Perseus Arm (magenta), and Outer Arm (red). The white dots indicate spurs or sources for which the arm assignment is unclear. The H{\sc I} emission is from the HI4PI survey~\citep{HI4PI2016}. The subfigure presents a zoomed-in view of the red box, in which the masers are assigned to the Norma Arm (orange crosses) and the Scutum Arm (orange triangles) by R19, respectively, while here they are assigned to the same arm, i.e., the Norma Arm.}
	\label{fig:lvmap}
\end{figure*}

\begin{figure*}[htbp]
	\center
	\includegraphics[width=0.8\textwidth]{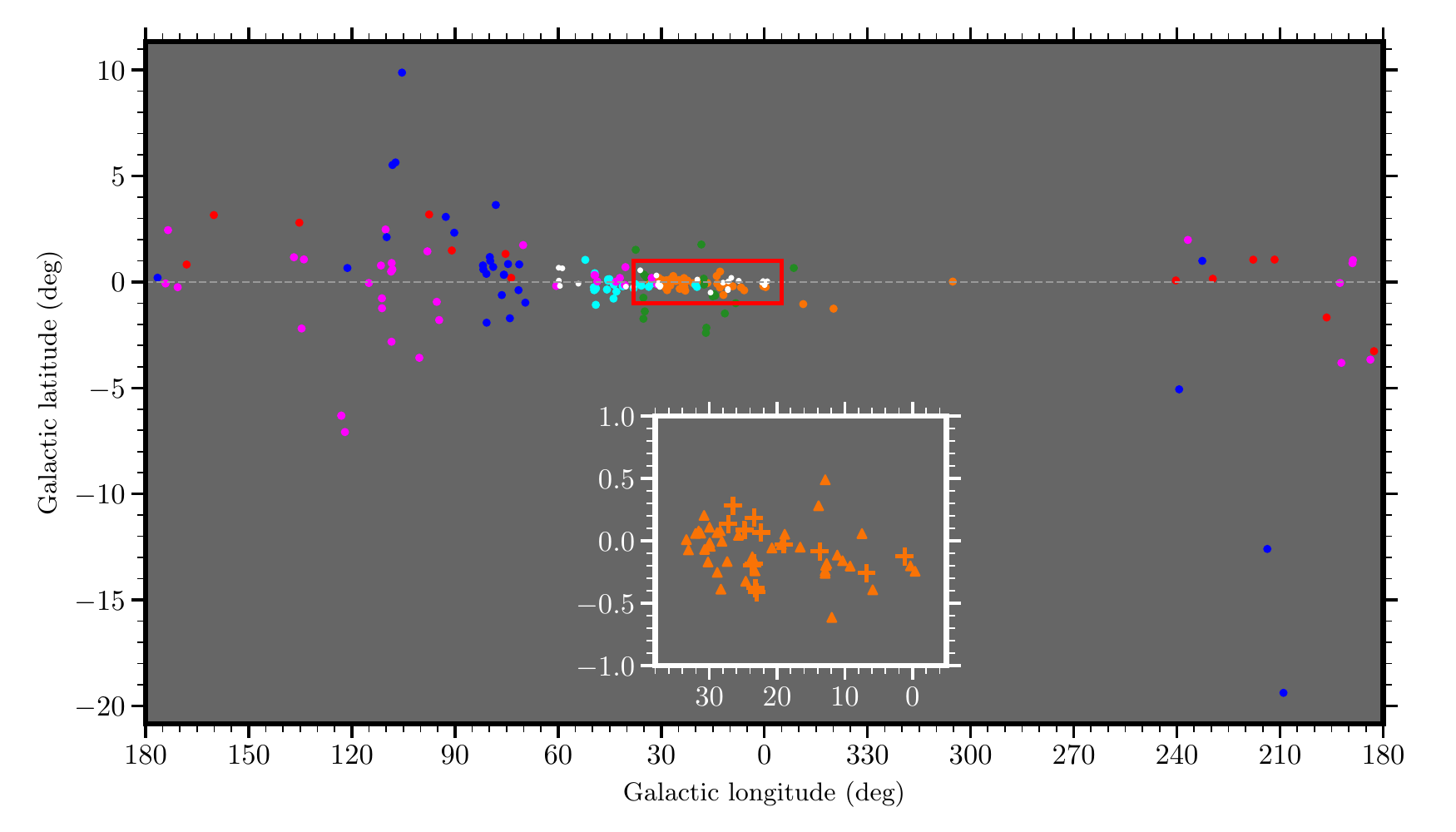}
	\caption{Galactic longitudes of the HMSFR masers as a function of Galactic latitude. The colored dots are maser sources with parallax measurements, same as in Figure~\ref{fig:lvmap}. The subfigure is the same as in Figure~\ref{fig:lvmap}.}
	\label{fig:lbmap}
\end{figure*}

\begin{figure*}[htbp]
	\center
	\includegraphics[width=0.8\textwidth]{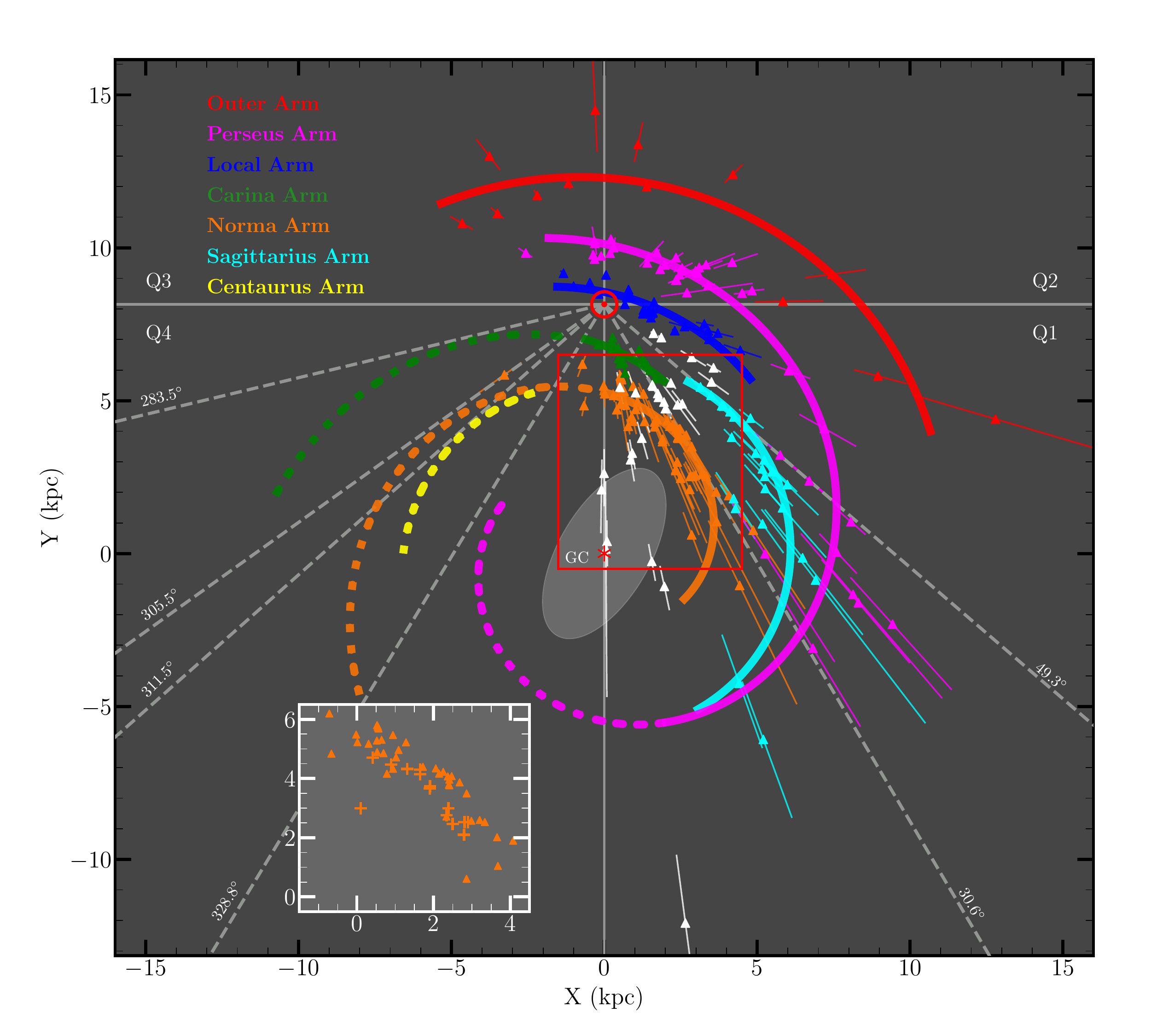}
	\caption{Plan view of the MW from the north Galactic pole showing the locations of the HMSFR masers (triangles) with measured trigonometric parallaxes and the spiral arms they trace. The distance uncertainties of the masers are indicated by error bars. The masers presented by white triangles indicate spurs or sources whose arm assignments are unclear. The spiral arms traced by masers are color coded solid lines and their names have been presented in the legend at the upper left. The dotted curved lines trace the centers of the inferred fitted spiral arms. The subfigure is the same as in Figure~\ref{fig:lvmap}. The Galactic bar is indicated by a shaded ellipse following \cite{hilmi2020}. The six spiral-arm tangencies are indicated by gray dashed lines. The GC (red asterisk) is at (0, 0) kpc and the Sun (red Sun symbol) is at (0, 8.15) kpc~\citep{reid2019}. Galactic rotation is clockwise.}
	\label{fig:xymap}
\end{figure*}

In Figure~\ref{fig:lvmap}, we have overlaid the HMSFR masers on the 
$l$--$v$ plot of H{\sc I} emission. Figure~\ref{fig:lbmap} shows the Galactic
longitude--latitude $(l, b)$ positions of the maser sources, and 
Figure~\ref{fig:xymap} presents a plan view of the MW. In these three figures, 
the white dots indicate spurs or sources for which the arm assignments are unclear.
The arm assignments for most sources, in terms of the $(l, v)$ matches, are 
clear and consistent with the parallax distances. It is worth mentioning that 
the maser sources assigned to the Norma and Scutum Arms in previous work 
(R19) are mixed together in the $l$--$v$, $l$--$b$, and $x$--$y$ plots (as 
shown by the zoomed-in subfigures in Figures~\ref{fig:lvmap}, \ref{fig:lbmap}, 
and \ref{fig:xymap}), respectively. Therefore, in this work, these sources are 
assigned to the same arm, namely the Norma Arm.
Besides, R19 indicated that the Sagittarius Arm has a 2 kpc long gap centered 
at $l \approx 43$$^{\circ}$ near ($x$, $y$) = (2, \,6) kpc (see 
Figure~\ref{fig:xymap}), and the arm segment on the left of the gap belongs to 
the Sagittarius Arm. As this arm segment is connected to the Carina Arm, here 
we have assigned the masers to the Carina Arm.

Since the radial velocities ($v$) of different spiral arms are almost equal to 0 
km s$^{-1}$ in the direction of the GC (340$^{\circ}$ $< l <$ 20$^{\circ}$) 
and anti-GC (160$^{\circ}$ $< l < $ 200$^{\circ}$), the Galactic latitudes and 
parallax distances are used as the primary indicators for arm assignment.
For sources within 5 kpc of the Sun, their arm assignments can be determined 
confidently based on their accurate parallax distances.
However, for some distant sources, their parallax distances may be consistent 
with two or even three spiral arms within their 1$\sigma$ uncertainties, so 
their Galactic latitudes can provide strong evidence to resolve this issue. 
G019.60$-$00.23 is a good example of this situation. Around $l \approx 
20$$^{\circ}$, $b$ and $v$ of the Sagittarius and Perseus Arms are 
($-$0.08$^{\circ}$, 39 km s$^{-1}$) and (0.08$^{\circ}$, 26 km s$^{-1}$), 
respectively~\citep{reid2016}. Near ($x$, $y$) = (4, $-$4) kpc, G019.60$-$00.23 
has a velocity of $\sim$41 km s$^{-1}$, implying it may belong to either the 
Sagittarius Arm or the Perseus Arm judging from the $(l, v)$ match in 
Figure~\ref{fig:lvmap} and the ($x$, $y$) match in Figure~\ref{fig:xymap}. 
Considering that the Galactic latitude of G019.60$-$00.23 is 
$-$0.23$^{\circ}$, we rule out assigning this maser to the Perseus Arm based 
on the latitude difference of 0.31$^{\circ}$ ($Z$-height of 70 pc at a distance 
of 13 kpc). The Sagittarius Arm matches this maser in both latitude and $v$, 
and thus we assign G019.60$-$00.23 to the Sagittarius Arm. The same 
technique for arm assignment is also adopted for the sources G229.57$+$00.15 
and G240.31$+$00.07, whose Galactic latitudes are 0.15$^{\circ}$ and 0.07$^{\circ}$, 
respectively. From the $l$--$v$ plot (Figure~\ref{fig:lvmap}) and the plan view of 
the MW (Figure~\ref{fig:xymap}), they could be in either the Perseus Arm or the 
Outer Arm. At $l \approx 230$$^{\circ}$ and  $\approx 240$$^{\circ}$, the 
Galactic latitudes of the Perseus Arm are $-$2.10$^{\circ}$ and 
$-$2.44$^{\circ}$~\citep{reid2019}, and the corresponding values are 
0.17$^{\circ}$ and 0.19$^{\circ}$ for the Outer  Arm~\citep{reid2019}.
Hence, the differences of 2.25$^{\circ}$ and 2.51$^{\circ}$ ($Z$-height of 
$\sim$200 pc at a distance of 5 kpc) in the Galactic latitude rule out the 
sources as belonging to the Perseus Arm, and instead they likely reside in 
the Outer Arm.
The arm assignment of each source is listed in Table~\ref{tab:masers}, and 
the numbers of the sources in the spiral arms are listed in Table~\ref{tab:arm_fit}.

\subsection{Young Objects}
\begin{figure*}[htbp]
    \centering
    \includegraphics[width=0.80\textwidth]{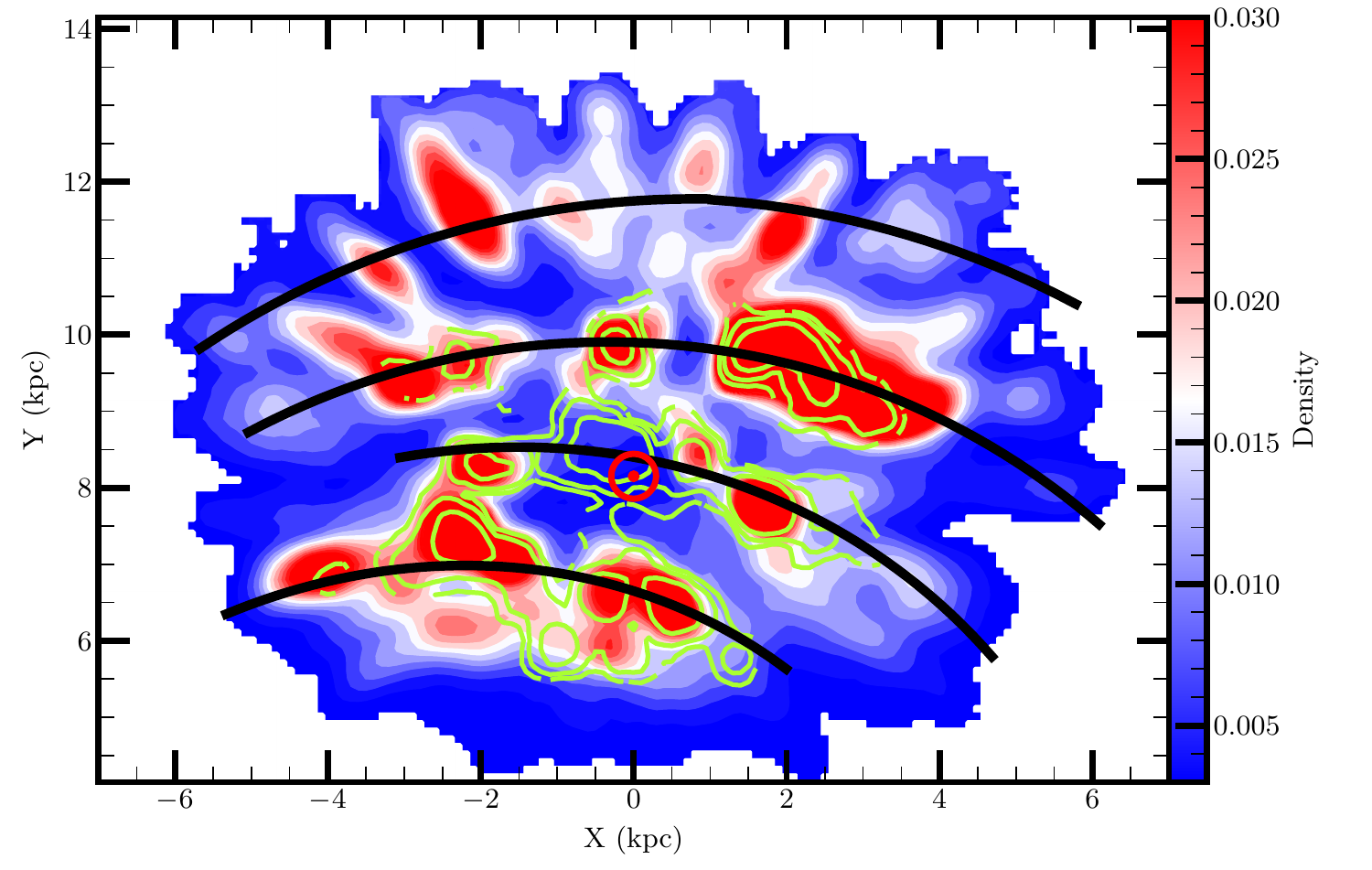}
    \caption{Density map of the O--B2-type stars and contours of the YOCs, which are over plotted with the MW's spiral arms~(the black lines) obtained by fitting for the positions of the O--B2-type stars. The red Sun symbol is the position of the Sun. For the original projected distributions of O--B2 stars and YOCs onto the Galactic disk, please see Figure~\ref{fig_ob_oc} in Appendix~\ref{af}.}
    \label{fig:OB_arm}
\end{figure*}

The high-precision parallaxes of young stars provided by \emph{Gaia} are 
excellent tracers, and can be used to map the spiral structure near the Sun. 
To reveal the MW's spiral structure better, we use a bivariate kernel density 
estimator~\citep{feigelson2012} to outline the stellar density structures 
around the Sun. Meanwhile, considering the exponential distribution of stars 
in the Galactic disk, we adopt a weight related to the Galactocentric radius, 
which is set as $w_a(R) = C \exp(R/R_d)$. Here, $C$ is the constant factor 
and $R_d$ is the disk scale, assumed to be 3 kpc \citep{binney2008}. 
Assuming that the weight at solar circle is 1, we simply set $C = \exp(-8.15/R_d)$.
Adopting a local bandwidth of 0.4 kpc, which is similar  to~\citet{poggio2021}, 
the density structures indicated by the O--B2-type stars (color background) 
and YOCs (contours) are shown in Figure~\ref{fig:OB_arm}. The color backgrounds 
and contours are concordant. The distributions of the O--B2-type 
stars and YOCs are highly structured rather than homogeneous. 

Owing to a lack of radial velocities of many O--B2-type stars, we could not 
assign them to spiral arms based on the Galactic $l$--$v$ maps as masers. 
Instead, we adopt a new method to judge which spiral arm an O--B2-type 
star belongs to according to the probability of the O--B2-type star being 
located in one of the spiral arms shown in the plan view of the MW. Here, 
the probability is calculated by a Gaussian function, i.e.,
\begin{equation}
\displaystyle P = \exp \frac{-(\Delta d)^{2}}{{2 \times} \max (\sigma_{a}^2, \, \sigma_{x}^2) },
\end{equation}
where $\Delta d$ is the distance to the spiral arm, $\sigma_{x}$ is the uncertainty 
of the distance, and $\sigma_{a}$ is the arm width, which defaults to 0.3~kpc in 
this work. Finally, we only select stars with probabilities larger than 10\%, 
corresponding to them being less than 1.7$\sigma$ away from the nearest spiral arm.

\subsection{Spiral-arm-fitting Function}
We fit a spiral pattern to the spiral-arm segments by adopting a log-periodic spiral, 
defined as,
\begin{equation}
    \ln (R / R_{\rm ref}) = -(\beta - \beta_{\rm ref}) \tan \psi,
    \label{eq:arms}
\end{equation}
where $R$ is the Galactocentric radius at the Galactocentric azimuth, $\beta$, 
(defined as 0 toward the Sun and increasing in the direction of Galactic rotation) 
for an arm with radius $R_{\rm ref}$ at reference azimuth $\beta_{\rm ref}$, 
and pitch angle $\psi$. We fit a straight line to $(x, y) = (\beta, \ln (R / R_{\rm ref}))$ 
using a Bayesian Markov Chain Monte Carlo (MCMC) procedure to estimate the 
parameters $R_{\rm ref}$ and $\psi$.

The $\chi^2$ test is employed to evaluate the fitted result of the spiral arms, i.e.,
\begin{equation}
   \left. \chi^{2} = \sum_{i} \left[\frac{(R_{{\rm mod}, i} - R_{{\rm obs}, i})^{2}}{\sigma^{2}} \right] \middle/ (N - 2).\right.
\end{equation}
Here, the subscripts ``obs" and ``mod" on the $R$, respectively mean the radius 
gleaned from the observational data and from the model, respectively; the subscript 
$i$ represents 
the $i$th source, $\sigma$ is the radius dispersion for $(R_{\rm mod} - 
R_{\rm obs})$, and $N$ is the number of sources.
Each log-periodic spiral arm has been depicted by the locations of the masers 
using the method described by R19. We also constrain the fitted arms to pass 
the observed arm tangencies~\citep{hou2015}.
After assigning the O--B2-type stars to the spiral arms in the solar neighborhood 
(the Outer, Perseus, Local, and Carina Arms) based on their probabilities of being
located in the referenced spiral arm, we fit the arm parameters ($R_{\rm ref}$ and 
$\psi$) using Eq.~\ref{eq:arms}. The referenced azimuth, $\beta_{\rm ref}$, is 
arbitrarily set to zero when using the O--B2-type stars to fit the spiral arms. 
Since the structures indicated by the O--B2-type stars seem to be consistent 
with those inferred by the masers, we use the spiral arms traced by the masers 
as the initial reference. Afterward, we assign the O--B2-type stars to the fitted 
arms and repeat the above procedures until the fitted spiral arms converge. 
The best-fitting parameters of the masers and O--B2-type stars are listed in 
Table~\ref{tab:arm_fit}. Figure~\ref{fig:xymap} presents the spiral arms depicted 
by the masers as well as the arm tangencies, and Figure~\ref{fig:OB_arm} shows 
the spiral arms depicted by the O--B2-type stars.

\section{The Morphology of the MW}\label{sec:results}
\subsection{The MW as a Multiple-arm Galaxy}
\label{multi-arm}

\begin{figure*}
    \centering
    \includegraphics[width=0.8\textwidth]{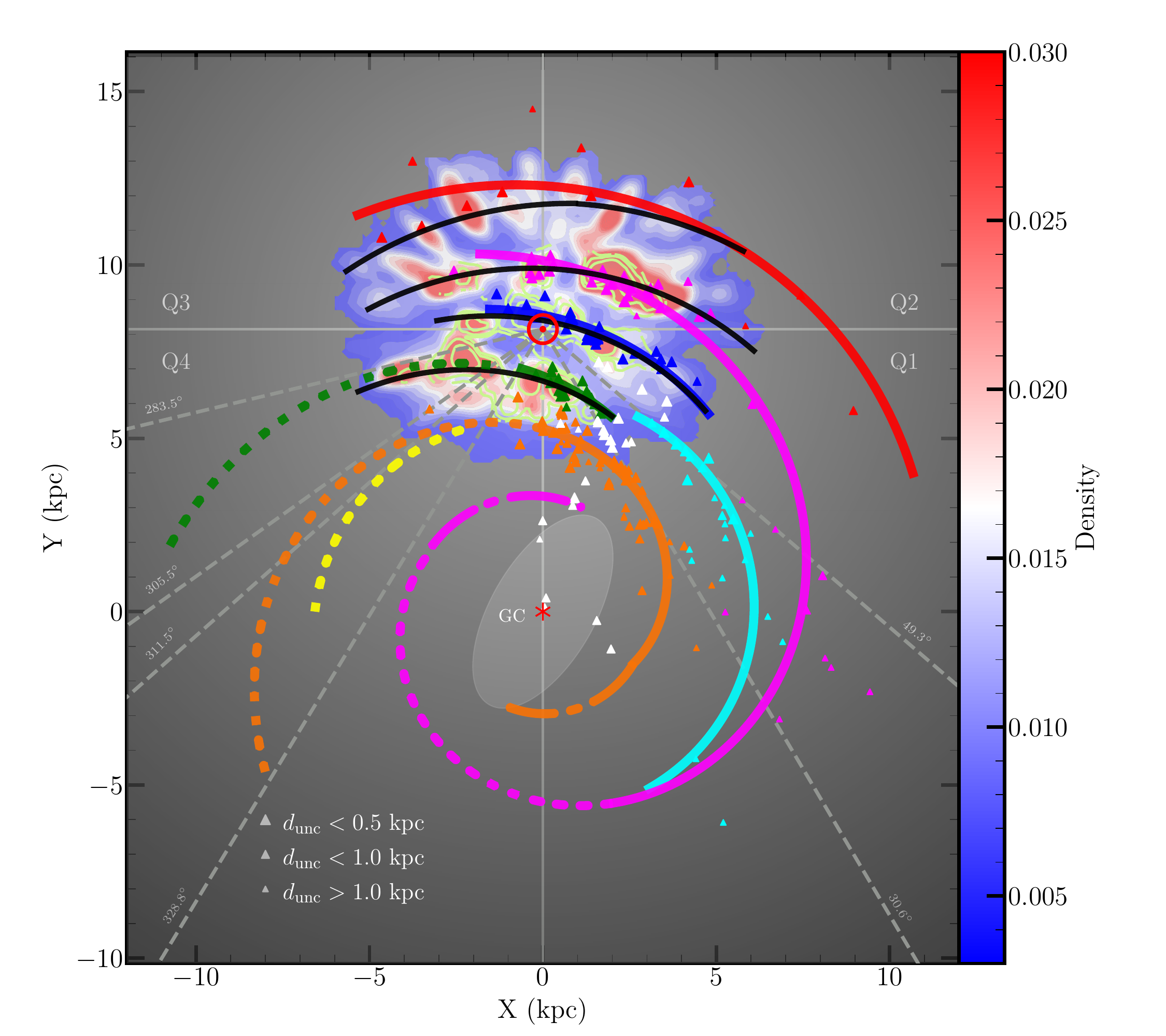}
    \caption{Plan view of the MW from the north Galactic pole showing the
    spiral arms traced by HMSFR masers (triangles), O--B2-type stars (color background), and YOCs (green contour). The distance uncertainties of the masers are indicated by the inverse size of the symbols, as given in the legend at the lower left. The other elements are the same as in Figure~\ref{fig:xymap}.
    }
    \label{fig:fig1}
  \end{figure*}

%
The precise locations of the masers, O--B2-type stars, and YOCs have revealed 
the spiral structure of the MW.
As shown in Figure~\ref{fig:fig1}, the Norma Arm traced by these masers
passes the arm tangency at $l$ $\sim$ 30.6$^{\circ}$ in the first Galactic quadrant.
Due to a lack of masers and young stars in the fourth Galactic quadrant, we adopt 
a constant pitch angle to extend the Norma Arm outward and it passes through the 
arm tangency at $l$ $\sim$ 305.5$^{\circ}$; the conjectural arm is depicted by the 
dotted line in Figure~\ref{fig:fig1}.
There is an additional arm tangency at $l$ $\sim$ 311.5$^{\circ}$ near the 
tangency of the Norma Arm (305.5$^{\circ}$).
Two adjacent tangencies are likely created by one spiral arm dividing into two spiral 
arms~\citep{minniti2021}.
Therefore, near the direction of $l$ $\sim$ 310$^{\circ}$--340$^{\circ}$, the Norma 
Arm likely bifurcates out of the Centaurus Arm.
Owing to a lack of high-precision data, the Centaurus Arm has been inferred by 
using the arm tangency at $l$ $\sim$ 311.5$^{\circ}$, as depicted by the dotted line 
in Figure~\ref{fig:fig1}.
After extending the Perseus Arm inward with a constant pitch angle, it naturally meets 
the arm tangency at $l$ $\sim$ 328.8$^{\circ}$, which is symmetric with the Norma 
Arm, as shown by the dotted line in Figure~\ref{fig:fig1}.
Employing the best-fitting pitch angle~(see Table~\ref{tab:arm_fit}), the Sagittarius 
Arm is expected to intersect with the Perseus Arm in the direction of $l$ $\sim$ 
10$^{\circ}$--30$^{\circ}$ and extend upwards passing the arm tangency at 
$l$ $\sim$ 49.3$^{\circ}$.
Around $l$ $\sim$ 10$^{\circ}$--30$^{\circ}$, in fact, the masers in the Sagittarius 
and Perseus Arms exhibit consistent trend in the $l$--$b$ and $l$--$v$ plots, which 
could provide support for the above intersection.
To sum up, the Norma and Perseus Arms are likely the two symmetric arms in the 
inner MW. As they extend from the inner Galaxy to the outer parts, they bifurcate, 
and connect to the Centaurus and Sagittarius Arms, respectively.

In addition to the Perseus Arm, the masers also trace several other spiral arms in the 
outer Galaxy region, including the Outer, Local, Carina, and Sagittarius Arms.
The Outer, Perseus, Local, and Carina Arms are confirmed by O--B2-type stars and 
YOCs, which also densify and complement the spiral arms depicted by the masers.
In Figure~\ref{fig:fig1}, the spiral arms depicted by the O--B2-type stars have been 
indicated by black solid lines.
The Local Arm has a propensity to bend down in the third quadrant and may extend 
forward. Due to a lack of data, we are currently unable to determine whether the 
Outer Arm is connected to one of the inner arms.
As shown in Figure~\ref{fig:fig1}, the Carina Arm may be not connected to 
the Sagittarius Arm.
Using masers and the arm tangency at $l$ $\sim$ 283.5$^{\circ}$, we have depicted 
the Carina Arm and inferred its structure in the fourth quadrant, which is confirmed 
by the fitted result of the young objects in the {\emph Gaia} data.
On the whole, the spiral structure in the solar neighborhood traced by the multifarious 
young objects nearly agrees with the structure depicted by R19. Considering the inner 
two symmetric arms and several arm segments in the outer Galaxy regions, the MW 
should be considered as a multiple-arm spiral galaxy.  

Previous studies have indicated that there are slight offsets between the spiral arms 
traced by objects of different ages~\citep[e.g.,][]{vallee2014,hou2015}. Indeed, as 
shown in Figure~\ref{fig:fig1}, there is a similar situation between the spiral arms 
drawn by masers and those by young objects; however, these differences are not 
expected to have an effect on the overall Galactic morphology determined in this work.
The consistency of the spiral arms indicated by the various tracers also reinforces 
the validity of these arms.
According to previous research on external galaxies, the pitch angle of a 
long spiral arm can be variable~\citep{honig2015}.
Although a constant pitch angle is typically adopted to depict the Perseus Arm, if 
it is not constant, it is not likely to change the global structure of the MW when 
considering the widespread distribution of the masers related to the Perseus Arm.

\subsection{Comparing the Galactic Morphology Here with That of R19}
In the solar neighborhood, the multiple spiral arms of the MW depicted by the 
masers, O--B2 stars, and YOCs are almost concordant with the results outlined 
by R19, and the young objects observed by \emph{Gaia} densify and extend the 
spiral structure constructed by using the VLBI maser data alone. 
In the inner region of the Galaxy, R19 proposed that our Galaxy has four major 
spiral arms, namely the Norma, Scutum, Sagittarius, and Perseus Arms, while this 
work suggests that there are two major spiral arms, the Norma and Perseus Arms.
Since the HMSFR masers assigned to the Norma and Scutum Arms in R19
are mixed together, these sources are considered as belonging to the same arm, 
i.e., the Norma Arm.
After extending inward into the inner regions of the MW, the Sagittarius 
and Perseus Arms are expected to intersect in the direction of 
$l$ $\sim$ 10$^{\circ}$--30$^{\circ}$, instead of being two spiral arms separated 
from each other. 
Therefore, the Sagittarius Arm in this work is considered to be a bifurcation of the 
Perseus Arm.

\subsection{Properties of MW-like Galaxies}
\begin{figure*}[htbp]
    \centering
    \includegraphics[width=0.6\textwidth]{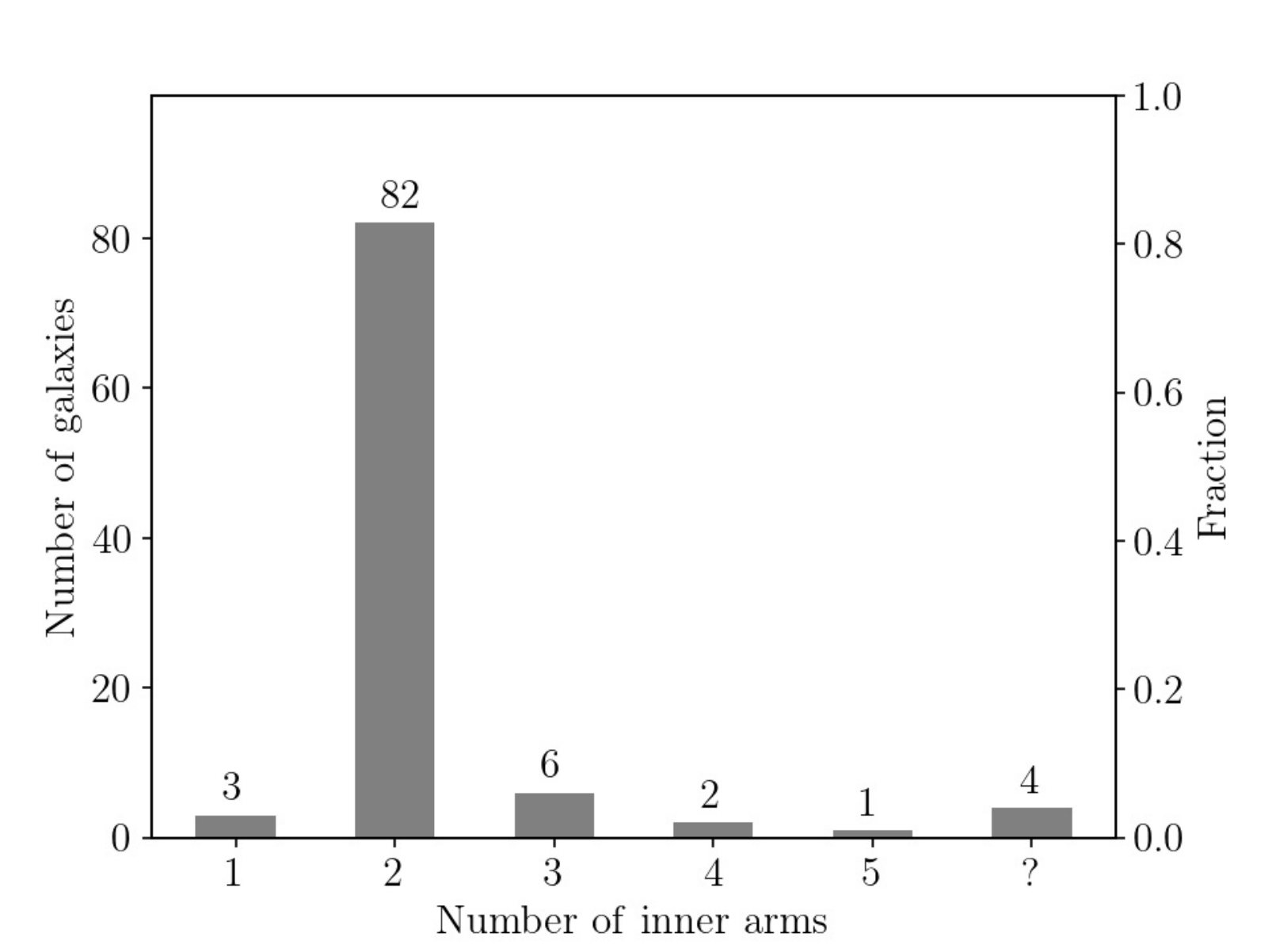}
    \caption{The number of inner spiral arms of multiple-arm galaxies in the sample.}
    \label{fig_inner}
\end{figure*}

Considering that the MW is a typical SBbc- or SBc-type galaxy~\citep{binney2008}, 
we select 185 external galaxies with morphologies identified as SBbc- and SBc-type 
from the SIMBAD database, and compare their morphologies with that of the MW. 
For more details, please see Appendix~\ref{extragalaxy}.
The properties of the sample of external MW-like galaxies, which contains 73 (40\%) 
flocculent, 99 (54\%) multiple-arm, 12 (6\%) grand-design, and one unidentified 
case, have been visually inspected.
For multiple-arm galaxies, we have also counted the number of their inner spiral 
arms, as shown in Figure~\ref{fig_inner}. Statistically, only 2\% of the multiple-arm 
galaxies have four inner arms. But relatively speaking, 83\% of the multiple-arm 
galaxies clearly present two inner arms.
As mentioned above, the MW galaxy likely has two inner symmetric arms and 
several arm segments in the outer regions, showing a multiple-arm morphology. So, in this case, 
the morphology of the MW is similar to those of most multiple-arm galaxies in the universe.

In most multiple-arm galaxies, there is usually a change in the appearance of the spirals 
midway out in the disk because of symmetric bifurcations or broadenings of the inner 
spiral arms.
Generally, the inner spiral arms of a galaxy are considered to be within 0.5~$R_{25}$ or 
twice the radius of galactic bar, where $R_{25}$ is the radius at which the surface 
brightness is 25 mag arcsec$^{-2}$~\citep{elmegreen1995}.
The $R_{25}$ value of the MW is about 11.5 kpc~\citep{deVaucouleurs1978}, and recent 
research has reported that the radius of the Galactic bar is $\sim$ 3.1 kpc~\citep{hilmi2020}.
Since the nearly symmetric bifurcations of the inner two arms of the MW are situated 
at distances of 5.0--6.5~kpc from the GC, the features of the inner two arms of our 
Galaxy are concordant with the statistical patterns of external galaxies.
Additionally, we find that the inner two arms of 73\% of the external multiple-arm, barred 
galaxies start from the ends of the galactic bars.
Because there are not enough young objects observed in the Galactocentric region, it is 
still not possible to determine whether or not the Norma and Perseus Arms start at the 
ends of the Galactic bar. The potential connections are depicted by chain lines in 
Figure~\ref{fig:fig1}.

\section{Summary}
\label{summary}

%
High-precision data of various young objects have suggested that the MW is a multiple-arm 
galaxy, which provides a possible alternative for future studies of the Galactic structure.
However, more details of the MW are expected to be revealed using VLBI parallax 
measurements in the southern sky and \emph{Gaia}'s upgraded data set.
Likewise, in the near future, the SKA and ngVLA will assist in obtaining maser measurements 
on the far side of the Galactic disk, allowing us to understand the spiral structure of our 
Galaxy more deeply.

\acknowledgments
We appreciate the anonymous referee for the instructive comments that helped us to improve the paper. This work was funded by the NSFC grant 11933011, National SKA Program of China (Grant No. 2022SKA0120103), and the Key Laboratory for Radio Astronomy. This work is based on observations made with the National Radio Astronomy Observatory's Very Long Baseline Array (VLBA), the Japanese VLBI Exploration of Radio Astrometry (VERA) project, the European VLBI Network (EVN), and the Australian Long Baseline Array (Australian LBA). This work has made use of data from the European Space Agency (ESA) mission \emph{Gaia} (\url{www.cosmos.esa.int/gaia}), processed by the Gaia Data Processing and Analysis Consortium (DPAC, \url{www.cosmos.esa.int/web/gaia/dpac/consortium}). This research has also made use of the SIMBAD database, operated at CDS, Strasbourg, France. Funding for the Sloan Digital Sky Survey (SDSS) and SDSS-II has been provided by the Alfred P. Sloan Foundation, the Participating Institutions, the National Science Foundation, the U.S. Department of Energy, the National Aeronautics and Space Administration, the Japanese Monbukagakusho, the Max Planck Society, and the Higher Education Funding Council for England. The SDSS Web Site is \url{http://www.sdss.org}.
The SDSS is managed by the Astrophysical Research Consortium for the Participating Institutions. The Participating Institutions are the American Museum of Natural History, Astrophysical Institute Potsdam, University of Basel, University of Cambridge, Case Western Reserve University, University of Chicago, Drexel University, Fermilab, the Institute for Advanced Study, the Japan Participation Group, Johns Hopkins University, the Joint Institute for Nuclear Astrophysics, the Kavli Institute for Particle Astrophysics and Cosmology, the Korean Scientist Group, the Chinese Academy of Sciences (LAMOST), Los Alamos National Laboratory, the Max-Planck-Institute for Astronomy (MPIA), the Max-Planck-Institute for Astrophysics (MPA), New Mexico State University, Ohio State University, University of Pittsburgh, University of Portsmouth, Princeton University, the United States Naval Observatory, and the University of Washington. The Digitized Sky Surveys were produced at the Space Telescope Science Institute under US Government grant NAG W-2166. The images of these surveys are based on photographic data obtained using the Oschin Schmidt Telescope on Palomar Mountain and the UK Schmidt Telescope. The plates were processed into the present compressed digital form with the permission of these institutions. The National Geographic Society Palomar Observatory Sky Atlas (POSS-I) was made by the California Institute of Technology with grants from the National Geographic Society.

\newpage

\begin{table*}[htp]
    \centering
    \caption{The Astrometric Parameters of the HMSFR Masers}
    \label{tab:masers}
     \setlength{\tabcolsep}{1.6mm}
     \renewcommand\arraystretch{1.2}
        \begin{tabular}{cccccccc}
            \hline \hline
Source & R.A. & Decl. & Parallax & $\mu$$_{x}$ & $\mu$$_{y}$ & $v$ & Spiral Arm\tnote{1} \\
 & (hh:mm:ss) & (dd:mm:ss) & (mas) & (mas yr$^{-1}$) & (mas yr$^{-1}$) & (km s$^{-1}$) &  \\ \hline
G305.20$+$00.01 & 13:11:16.8912 & $-$62:45:55.008 & 0.250 $\pm$ 0.050 & $-$6.90 $\pm$ 0.33 & $-$0.52 $\pm$ 0.33 & $-$38 $\pm$ 5 & Nor \\
G339.88$-$01.25 & 16:52:04.6776 & $-$46:08:34.404 & 0.480 $\pm$ 0.080 & $-$1.60 $\pm$ 0.52 & $-$1.90 $\pm$ 0.52  & $-$34 $\pm$ 3 & Nor \\
G348.70$-$01.04 & 17:20:04.0360 & $-$38:58:30.920 & 0.296 $\pm$ 0.026 & $-$0.73 $\pm$ 0.31 & $-$2.83 $\pm$ 0.59  & $-$9 $\pm$ 5 & Nor \\
G359.61$-$00.24 & 17:45:39.0697 & $-$29:23:30.265 & 0.375 $\pm$ 0.021 & 1.00 $\pm$ 0.40 & $-$1.50 $\pm$ 0.50 & 21 $\pm$ 5 & Nor \\
G007.47$+$00.05 & 18:02:13.1823 & $-$22:27:58.981 & 0.049 $\pm$ 0.006 & $-$2.43 $\pm$ 0.10 & $-$4.43 $\pm$ 0.16 & $-$14 $\pm$ 10 & Nor \\
G000.31$-$00.20 & 17:47:09.1092 & $-$28:46:16.278 & 0.342 $\pm$ 0.042 & 0.21 $\pm$ 0.39 & $-$1.76 $\pm$ 0.64 & 18 $\pm$ 3 & Nor \\
G005.88$-$00.39 & 18:00:30.2801 & $-$24:04:04.576 & 0.334 $\pm$ 0.020 & 0.18 $\pm$ 0.34 & $-$2.26 $\pm$ 0.34 & 9 $\pm$ 5 & Nor \\
G009.21$-$00.20 & 18:06:52.8421 & $-$21:04:27.878 & 0.303 $\pm$ 0.096 & $-$0.41 $\pm$ 0.45 & $-$1.69 $\pm$ 0.50 & 43 $\pm$ 5 & Nor \\
G010.32$-$00.15 & 18:09:01.4549 & $-$20:05:07.854 & 0.343 $\pm$ 0.035 & $-$1.03 $\pm$ 0.36 & $-$2.42 $\pm$ 0.50 & 10 $\pm$ 5 & Nor \\
G011.10$-$00.11 & 18:10:28.2470 & $-$19:22:30.216 & 0.246 $\pm$ 0.014 & $-$0.23 $\pm$ 0.38 & $-$2.01 $\pm$ 0.41 & 29 $\pm$ 5 & Nor \\
G011.91$-$00.61 & 18:13:58.1205 & $-$18:54:20.278 & 0.297 $\pm$ 0.050 & 0.66 $\pm$ 0.69 & $-$1.36 $\pm$ 0.75 & 37 $\pm$ 5 & Nor \\
G012.68$-$00.18 & 18:13:54.7457 & $-$18:01:46.588 & 0.416 $\pm$ 0.028 & $-$1.00 $\pm$ 0.95 & $-$2.85 $\pm$ 0.95 & 58 $\pm$ 10 & Nor \\
 & & &  ...\\
            \hline
        \end{tabular}
        \tablecomments{Nor, Norma Arm; Sag, Sagittarius Arm; Loc, Local Arm; Per, Perseus Arm; Out, Outer Arm; Car, Carina Arm. Sources indicated with ``???'' could not be confidently assigned to an arm. $v$: LSR velocity.}
\end{table*}

\begin{table*}[htp]
    \centering
    \caption{Spiral-arm Parameters}
    \label{tab:arm_fit}
    \setlength{\tabcolsep}{1.0mm}
    \renewcommand\arraystretch{1.2}
        \begin{tabular}{lcrrrrrccr}
        \hline \hline
Spiral Arm & $N$ & $\beta$ Range & $R_{\rm ref}$  & $\beta_{\rm ref}$  & $\psi$   & Width   & $l$ Tangency & $\chi^{2}$  &   Tracer   \\
  &  &  (deg) &  (kpc) &  (deg) &  (deg) &  (kpc) & (deg) &   &     \\ \hline
Outer & 12 &  $-$25 $\rightarrow$ 70   & 12.05 $\pm$ 0.39 & 17.9 $\pm$ 1.94 & 3.6 $\pm$ 4.1 & 0.96 $\pm$ 0.18 & -- & 1.08 & masers\\
 & 231 & $-$30 $\rightarrow$ 5 & 11.75 $\pm$ 0.02 & 0 $\pm$ (--) & $-$3.9 $\pm$ 0.5 & 0.41 $\pm$ (--) & -- & 1.08 & O--B2 stars\\
  & 626 & 5 $\rightarrow$ 30 & 11.79 $\pm$ 0.05 & 0 $\pm$ (--) & $-$1.1 $\pm$ 1.4 & 0.41 $\pm$ (--) & -- & 1.02 & O--B2 stars\\
Perseus & 39 & $-$10 $\rightarrow$ 160 & 9.50 $\pm$ 0.09 & 18.78 $\pm$ 2.26 & 11.0 $\pm$ 1.1 & 0.41 $\pm$ 0.10 & -- & 1.00 & masers \\
 & 5\,797 & $-$30 $\rightarrow$ 40 & 9.90 $\pm$ 0.01 & 0 $\pm$ (--) & 1.9 $\pm$ 0.3 & 0.28 $\pm$ (--) & -- & 1.00 & O--B2 stars \\
Local & 28  & $-$10 $\rightarrow$ 40 & 8.29 $\pm$ 0.08 & 8.95 $\pm$ 2.12 & 11.3 $\pm$ 1.9 & 0.30 $\pm$ 0.03 & -- & 1.04  & masers \\
 & 2\,363  & $-$20 $\rightarrow$ 40 & 8.40 $\pm$ 0.16 & 0 $\pm$ (--) & 10.0 $\pm$ 2.5 & 0.26 $\pm$ (--) & -- & 1.09  & O--B2 stars \\
Carina & 17 & $-$5 $\rightarrow$ 20 & 6.35 $\pm$ 0.11 & 9.93 $\pm$ 1.88 & 21.4 $\pm$ 4.6 & 0.29 $\pm$ 0.06 & -- & 1.00  & masers \\
 & 6\,330 & $-$40 $\rightarrow$ 20 & 6.65 $\pm$ 0.02 & 0 $\pm$ (--) & 17.5 $\pm$ 0.9 & 0.26 $\pm$ (--) & -- & 1.06  & O--B2 stars \\
Sagittarius & 24 & 26 $\rightarrow$ 150 & 6.02 $\pm$ 0.65 & 119.11 $\pm$ 2.03 & 1.3 $\pm$ 5.1 & 0.47 $\pm$ 0.60 & 49.3 & 1.03 & masers \\
Norma & 53 & 0 $\rightarrow$ 120 & 4.96 $\pm$ 0.02 & 13.07 $\pm$ 2.79 & 15.0 $\pm$ 1.1 & 0.36 $\pm$ 0.06 & 30.6, 305.5 & 1.01 & masers \\
            \hline
\end{tabular}
\tablecomments{$N$: the number of HMSFR masers and O--B2-type stars used in the fitting.}
\end{table*}

\appendix

\setcounter{table}{0}
\setcounter{figure}{0}
\setcounter{equation}{0}
\renewcommand{\thetable}{A\arabic{table}}
\renewcommand{\thefigure}{A\arabic{figure}}
\renewcommand{\theequation}{A\arabic{equation}}

\section{Additional figures}
\label{af}
\begin{figure*}[htbp]
    \centering
    \subfigure[O--B2-type stars]{\includegraphics[width=0.49\textwidth]{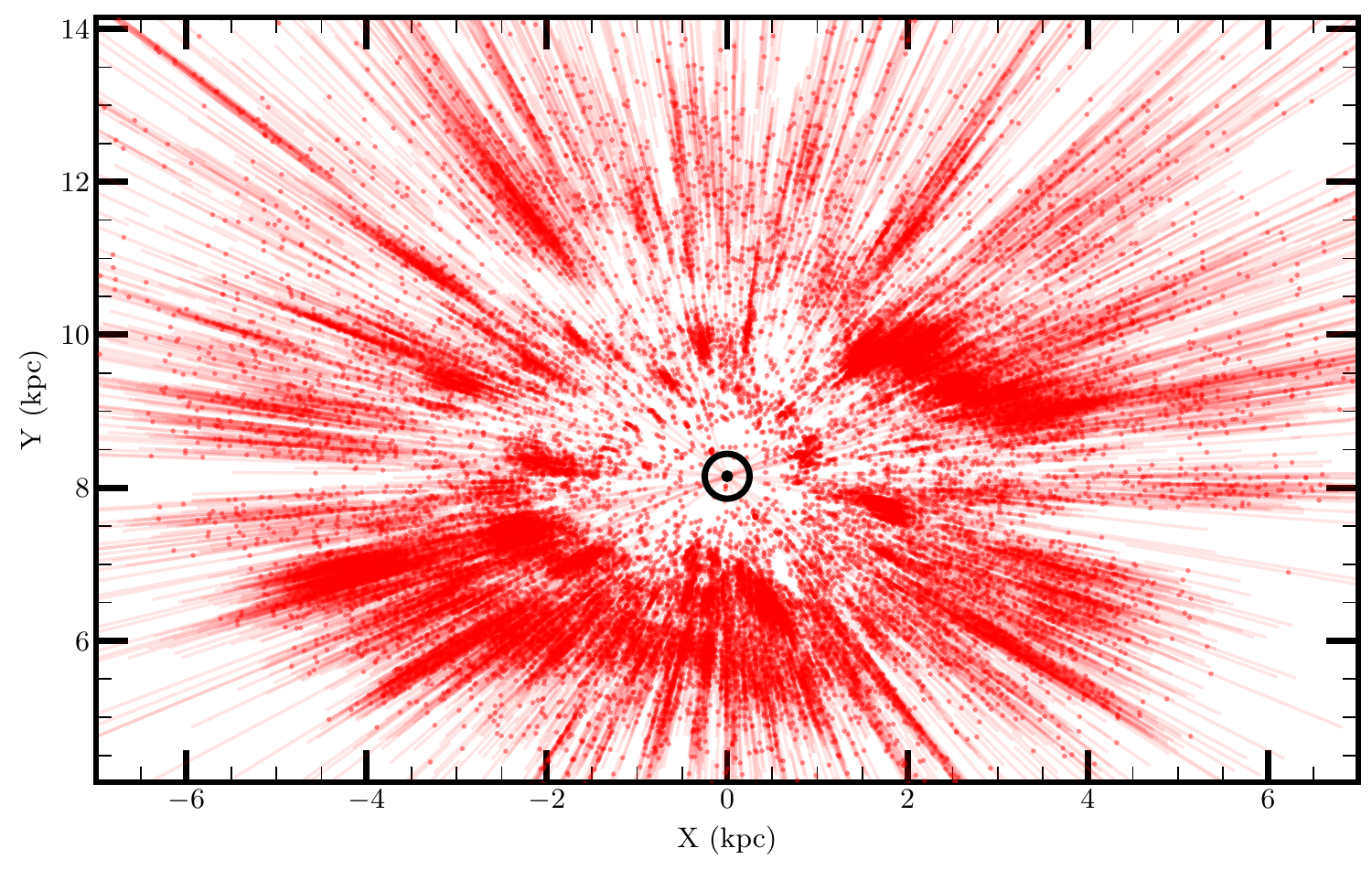}}
    \subfigure[YOCs]{\includegraphics[width=0.49\textwidth]{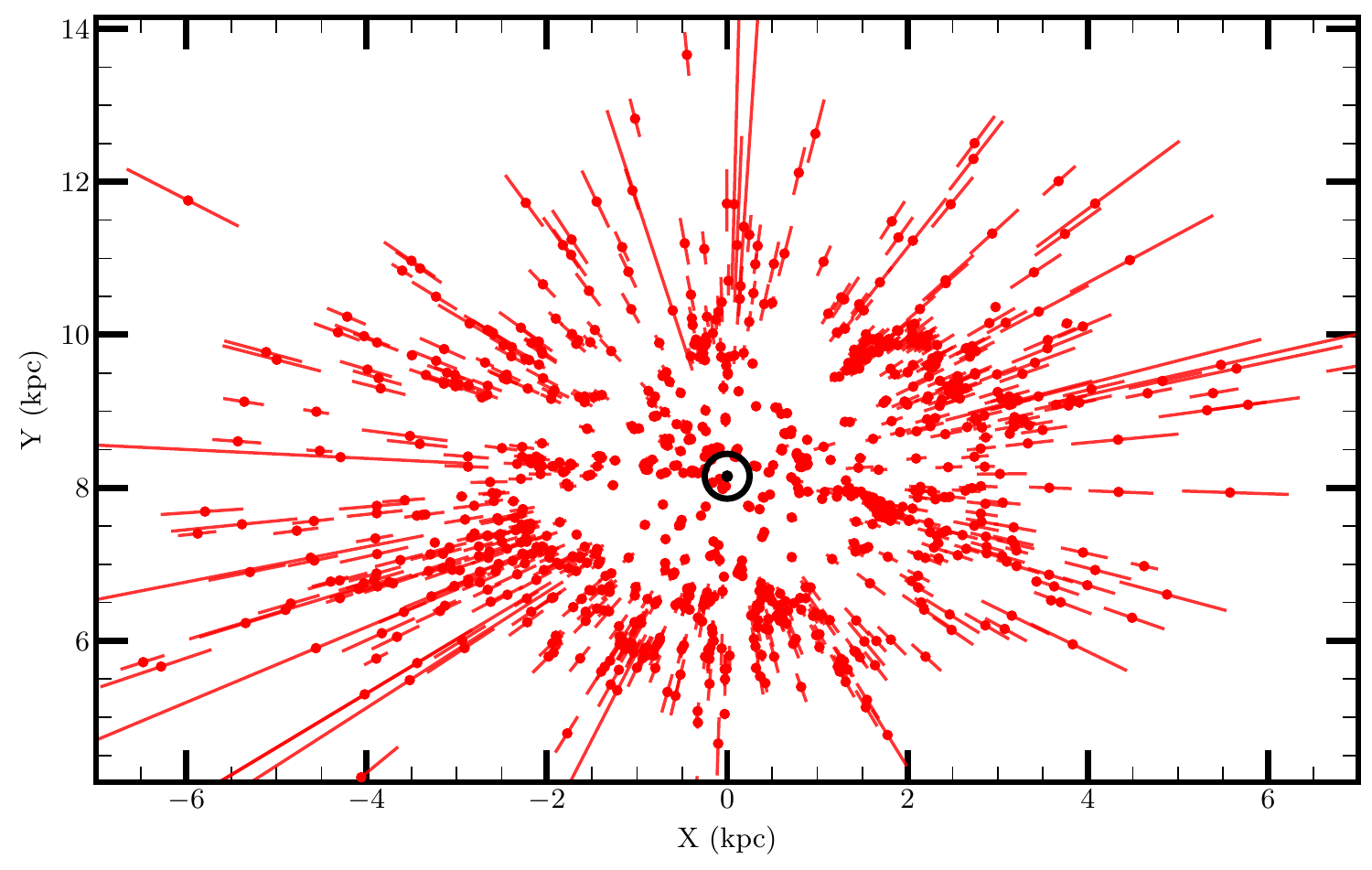}}
    \caption{Distributions of the O--B2-type stars (a) and YOCs (b) projected onto 
    the Galactic disk, together with their 1$\sigma$ distance uncertainties.
    }
    \label{fig_ob_oc}
\end{figure*}

Figure~\ref{fig_ob_oc} displays the distributions of the O--B2 stars and YOCs 
projected onto the Galactic plane.

\section{MW-like Galaxies}
\label{extragalaxy}

The contribution of the O--B-type stars to the total MW intensity is only 15\% in 
the near-infrared, compared with 65\% in bluer filters~\citep{schweizer1976}. 
In order to reveal the spiral morphology of young objects, we use blue images (defined 
below) in this work.
Here, the imaging data of the selected galaxies were obtained from the Sloan Digital 
Sky Survey (SDSS) DR7 in the $g$ band or Digitized Sky Survey (DSS) in the blue band.

\begin{figure*}[htbp]
    \centering
    \subfigure[ESO 443-69]{\includegraphics[width=0.32\textwidth]{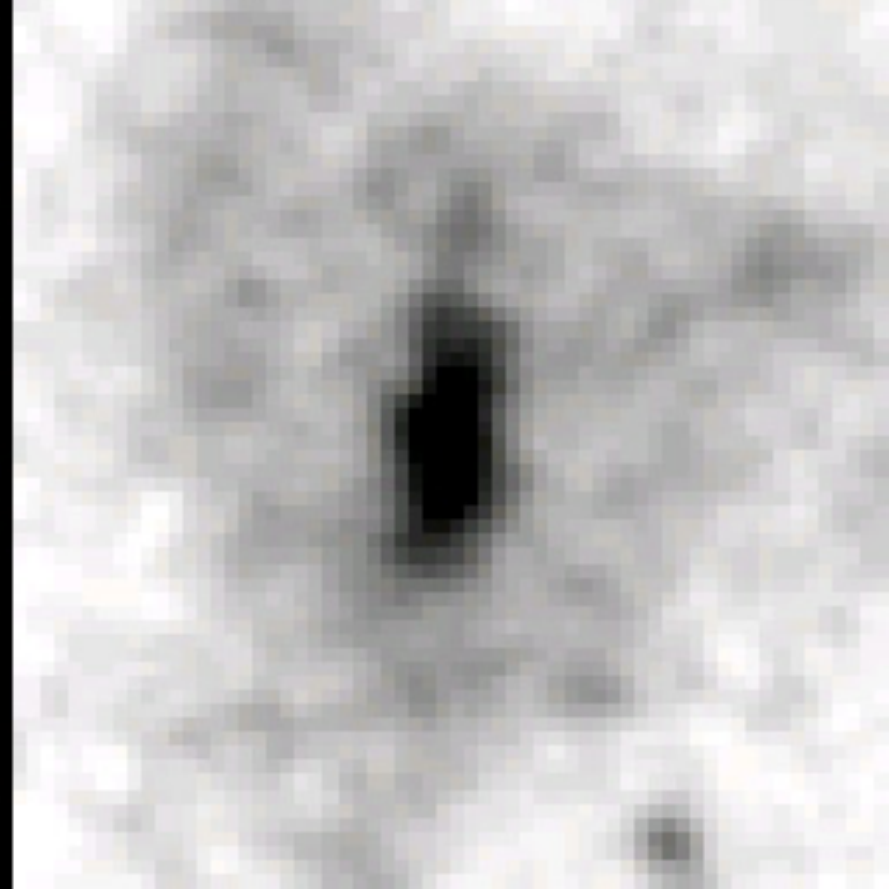}}
    \subfigure[ESO 533-45]{\includegraphics[width=0.32\textwidth]{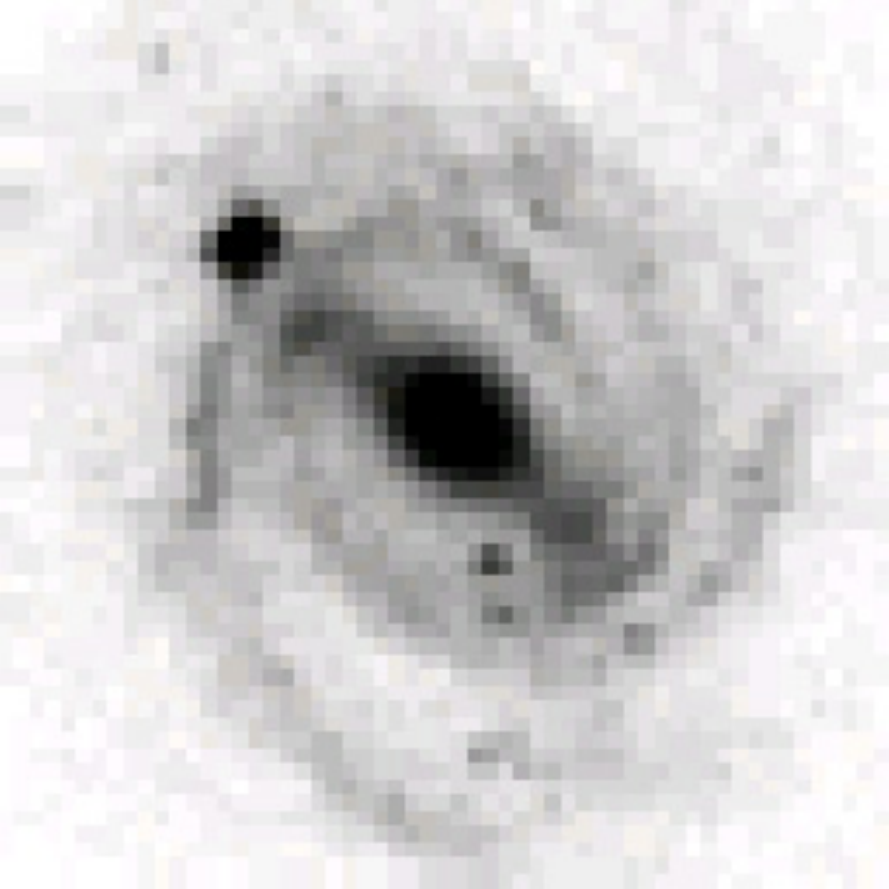}}
    \subfigure[IC  167]{\includegraphics[width=0.32\textwidth]{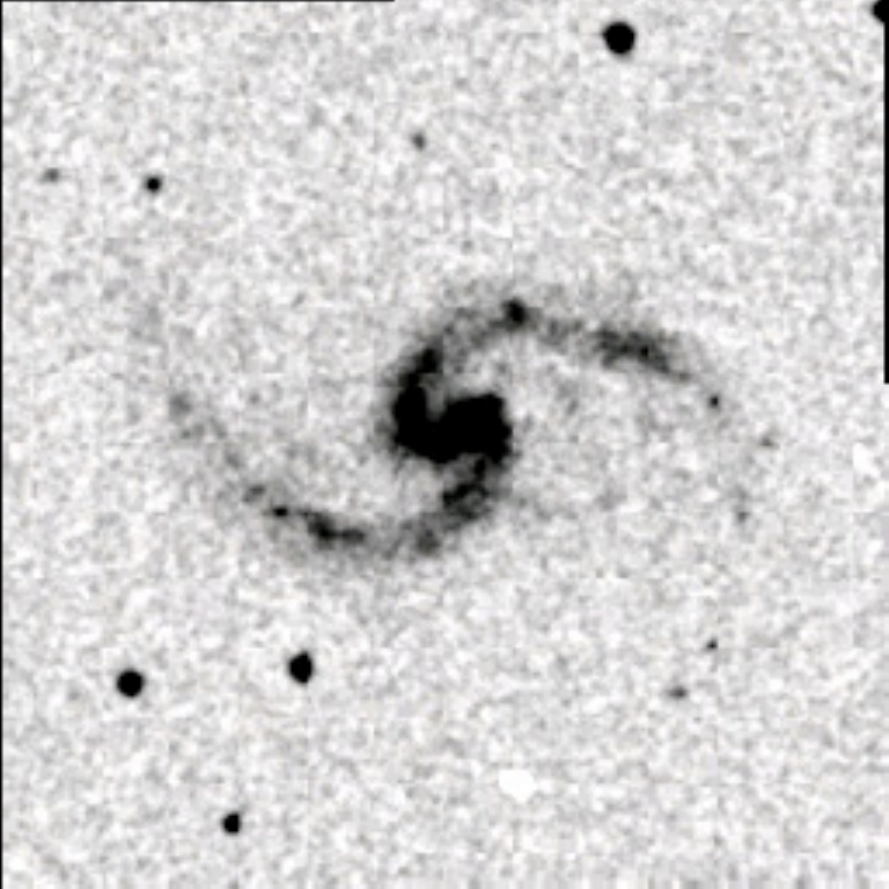}}
    \caption{Examples of flocculent (a), multiple-arm (b), and grand-design (c) galaxies.}
    \label{fig_sample_galaxy}
\end{figure*}

After excluding galaxies with small angular sizes (minor diameter smaller than 30 
arcsec) and large inclinations (inclinations larger than 60$^\circ$), we obtain a sample 
of 185 external MW-like galaxies. The spiral-arm classification method we adopt is 
from~\cite{elmegreen1987}. First, a flocculent galaxy lacks bimodal symmetry and 
has a spiral-like structure composed only of small pieces. Second, a multiple-arm 
galaxy generally has two inner arms branching midway out in the disk, and there 
are other arms as well. Third, a grand-design galaxy has a two-arm symmetry, 
and the arms are long and continuous at least over part of the galaxy.  We identify
the spiral-arm classifications of the selected 185 external MW-like galaxies, as listed in 
Table~\ref{tab:exgal}, where the flocculent, multiple-arm, and grand-design galaxies 
are symbolized as F, M, and G, respectively. Figure~\ref{fig_sample_galaxy} presents 
some examples of the flocculent, grand-design, and multiple-arm galaxies.

\begin{figure*}[htbp]
    \centering
    \subfigure[NGC 289]{\includegraphics[width=0.3\textwidth]{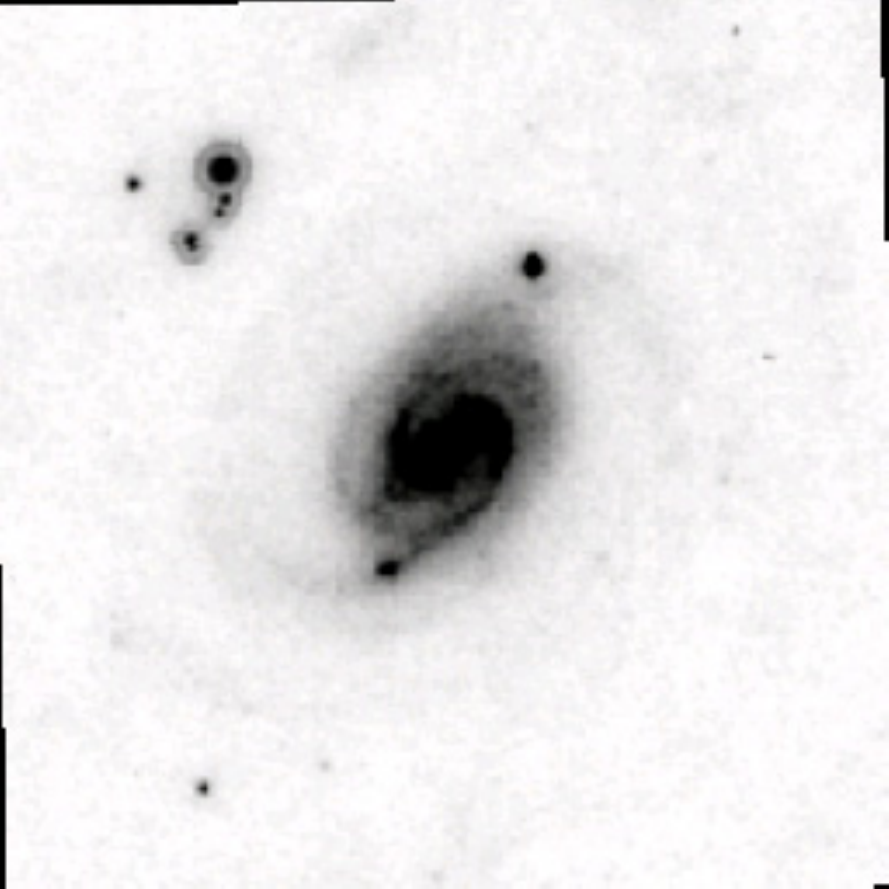}}
    \subfigure[NGC 1300]{\includegraphics[width=0.3\textwidth]{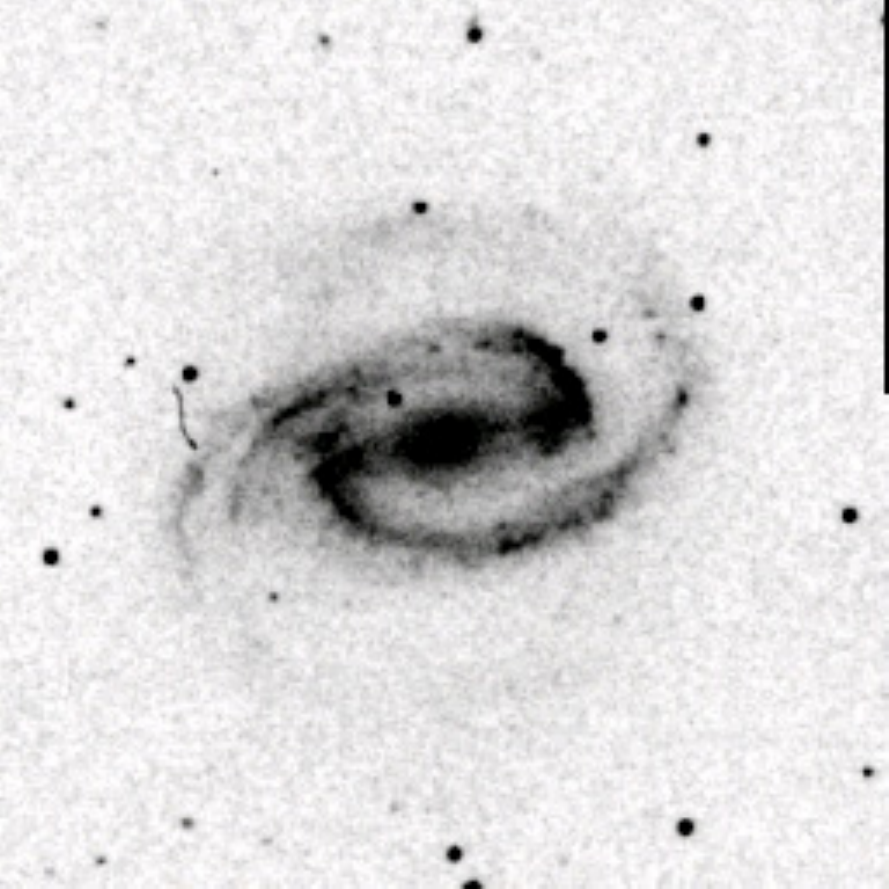}}
    \subfigure[NGC 2441]{\includegraphics[width=0.3\textwidth]{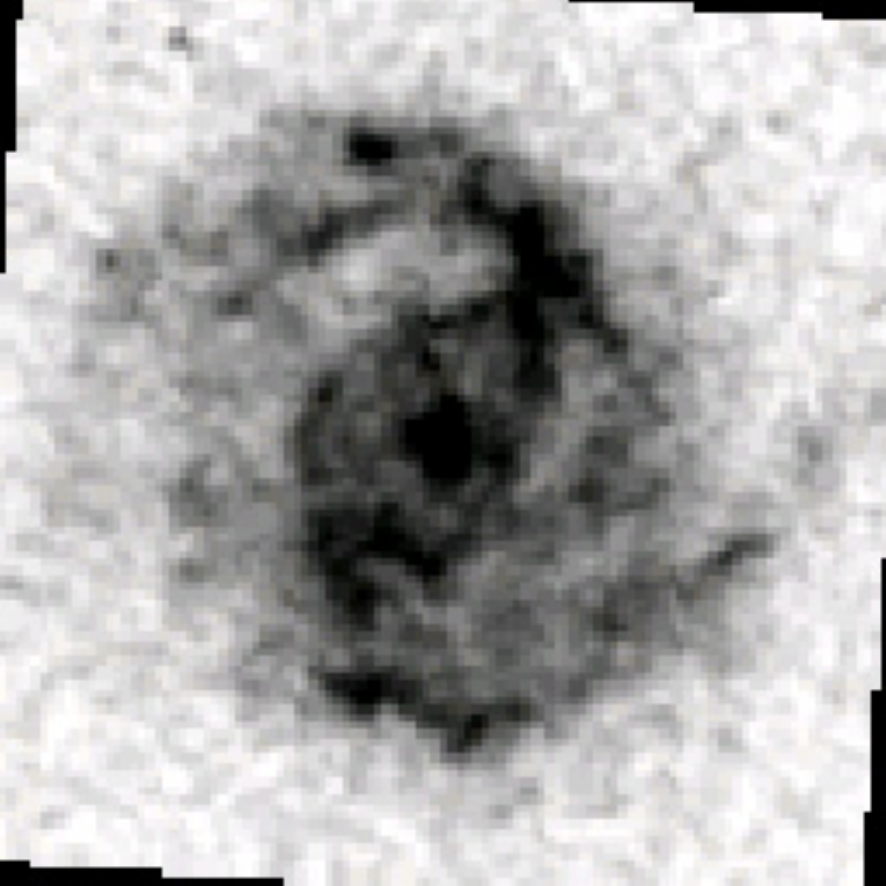}}
    \subfigure[NGC 3145]{\includegraphics[width=0.3\textwidth]{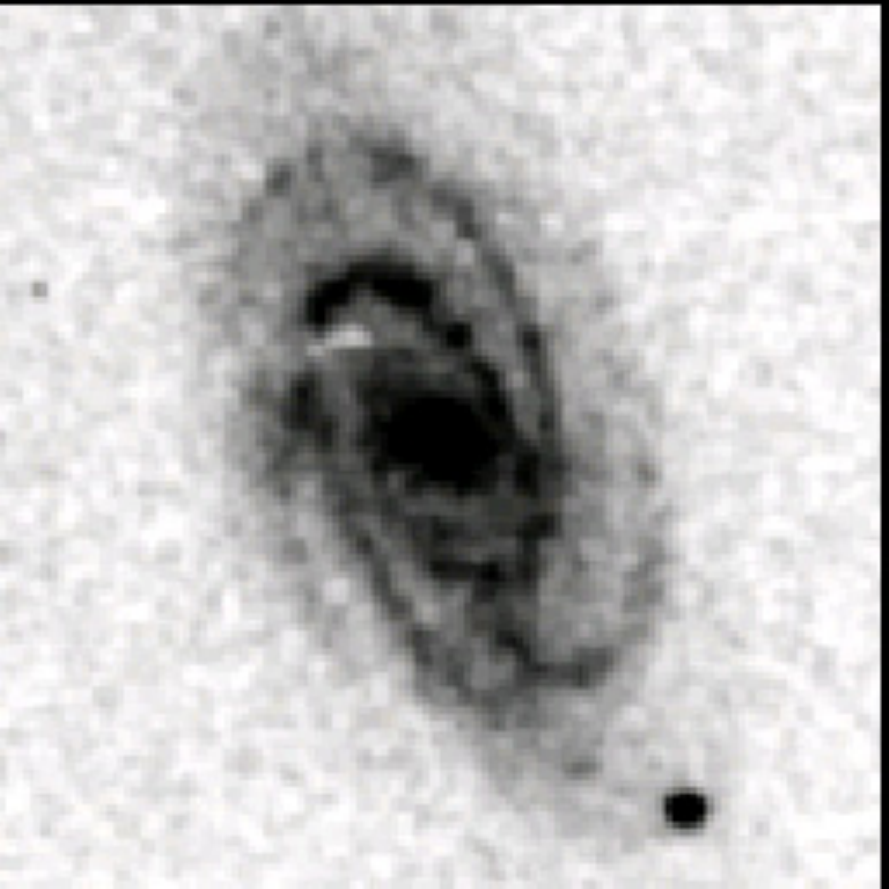}}
    \subfigure[NGC 3687]{\includegraphics[width=0.3\textwidth]{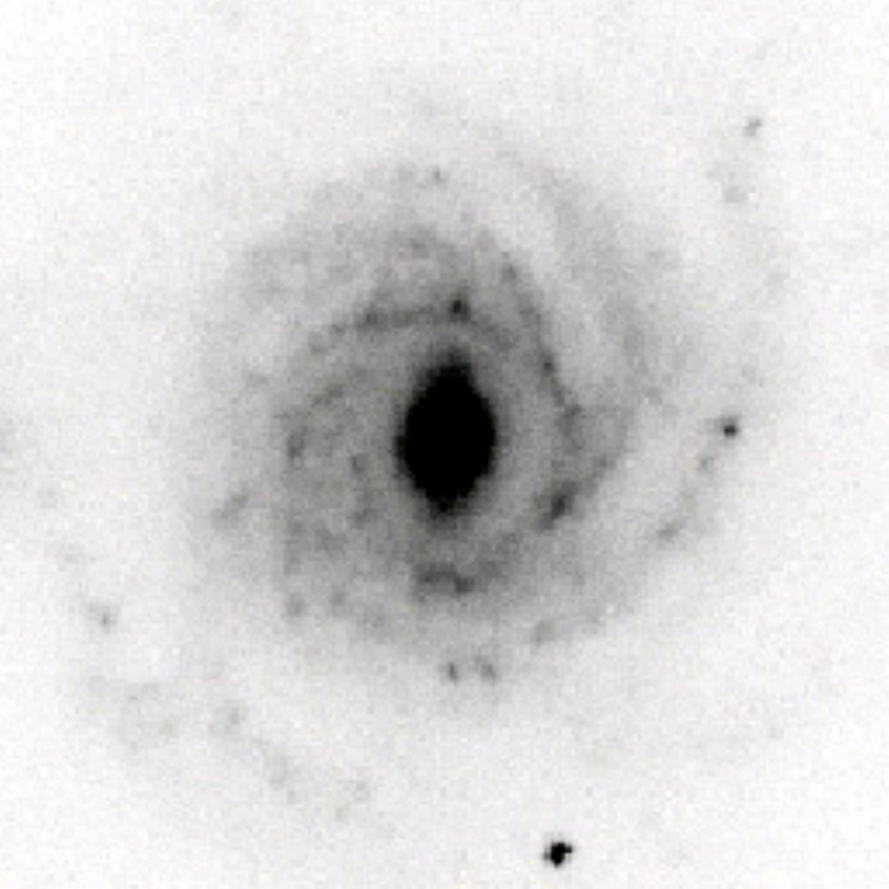}}
    \caption{Images of galaxies with different spiral-arm classifications in this work and~\cite{elmegreen1987}. (a) NGC 289 was classified as a grand-design galaxy by~\cite{elmegreen1987}, while we can clearly identify the bifurcation of an inner arm in the middle bottom in the galaxy. Thus, we classify NGC 289 as a multiple-arm galaxy. (b) NGC 1300 was classified as a grand-design galaxy by~\cite{elmegreen1987}, while the two spiral arms from the right side of the galactic bar are not continuous. Thus, we classify NGC 1300 as a multiple-arm galaxy, and this galaxy has inner three arms. (c) NGC 2441 clearly has several arms and is classified as a multiple-arm galaxy by us rather than a grand-design galaxy, as by~\cite{elmegreen1987}. (d) and (e) NGC 3145 and NGC 3687 also clearly have several arms and are classified as multiple-arm galaxies in this work rather than as flocculent galaxies, as by~\cite{elmegreen1987}.}
    \label{fig_differ_class}
\end{figure*}

Using the high-resolution $B$-band images provided by the Palomar Observatory 
Sky Survey, spiral-arm classifications were made for more than 700 external galaxies 
with various types by classified \citep{elmegreen1987}. 
Table~\ref{tab:compare} presents the classifications of 29 galaxies that appear in 
this work and in that of~\cite{elmegreen1987}, in which 24 galaxies have the same 
spiral-arm classes, and five galaxies are classified as multiple-arm galaxies in this 
work, as shown in Figure~\ref{fig_differ_class}. 
Three of the five galaxies, i.e., NGC 289, NGC 1300, and NGC 2441, which were 
classified as grand-design galaxies by~\cite{elmegreen1987}, tend to be multiple-arm 
galaxies due to their bifurcate or noncontinuous inner spiral arms. Several spiral 
arms of another two spiral galaxies, i.e., NGC 3145 and NGC 3687, also show that 
they are multiple-arm galaxies, rather than flocculent galaxies, as classified by~\cite{elmegreen1987}.

\begin{table*}[htp]
    \centering
    \caption{Properties of the External Galaxies}
    \label{tab:exgal}
    \setlength{\tabcolsep}{1.3mm}
    \renewcommand\arraystretch{1.15}
        \begin{tabular}{lcccccccc}
            \hline \hline
Name & Morphological Type &  $D_{maj}$  &  $D_{min}$ & Survey  & EE Type  & Inner Two Arms  & Arm Start From Bar & EE87 \\
 &  & (arcmin)  &  (arcmin)  &   &   & (Y/N)  & (Y/N) &  \\ \hline
ESO  60-18 &          SBc & 0.73 & 0.61 &  DSS & G &   Y &   Y & -- \\
ESO  61-6 &          SBc & 0.85 & 0.51 &  DSS & F & -- & -- & -- \\
ESO  69-9 &         SBc: & 1.86 & 1.32 &  DSS & M &   Y &   Y & -- \\
ESO  72-12 &          SBc & 1.15 & 0.65 &  DSS & M &   Y &   Y & -- \\
ESO  76-22 &          SBc & 0.93 & 0.68 &  DSS & M &   Y &   Y & -- \\
ESO  85-5 &          SBc & 1.15 & 0.71 &  DSS & M &   Y &   Y & -- \\
ESO 109-34 &          SBc & 1.05 & 0.68 &  DSS & M &   Y &   N & -- \\
ESO 111-15 &       SBc/Ir & 0.76 & 0.68 &  DSS & F & -- & -- & -- \\
ESO 142-4 &         SBc: & 0.95 & 0.60 &  DSS & G &   Y &   Y & -- \\
ESO 193-6 &          SBc & 1.23 & 1.05 &  DSS & M &   Y &   Y & -- \\
ESO 217-32 &          SBc & 1.07 & 0.79 &  DSS & G &   Y &   Y & -- \\
ESO 221-14 &    SB:c/I... & 0.95 & 0.76 &  DSS & M &   Y &   Y & -- \\
&&&&            ...\\
            \hline
        \end{tabular}
         \tablecomments{
         Column 1: Name of the external galaxy.
         Column 2: Morphological type from the SIMBAD database. The colons are used to indicate that the family and variety are uncertain interpretations in this case.
         Column 3: Major diameter of the galaxy from the SIMBAD database. The units are arcminutes.
         Column 4: Minor diameter of the galaxy from the SIMBAD database. The units are arcminutes.
         Column 5: Image data come from the DSS blue band or SDSS DR7 $g$ band.
         Column 6: Spiral-arm classes in this work. F, flocculent, M, multiple-arm, and G, grand-design.
         Column 7: Whether there are two inner arms or not. If not, the number of inner arms is presented in the parenthesis, and ``--'' represents that the number of inner arms is hard to identify.
         Column 8: Whether the inner spiral arms start from the end of bar or not. 
         Column 9: Arm classes in~\cite{elmegreen1987}. 1--4 represent flocculent galaxies, 5--9 represent multiple-arm galaxies, and 10--12 represent a grand-design galaxy.}
\end{table*}

\begin{table*}[htbp]
    \centering
    \setlength\tabcolsep{29pt}
    \renewcommand\arraystretch{2.0}
    \caption{Spiral-arm Classifications of 29 External Galaxies in This Work And~\cite{elmegreen1987}}
    \label{tab:compare}
    \begin{tabular}{c|ccc}
    \hline\hline
    \diagbox{This Work}{EE87} & F & M & G\\ 
    \hline
    F & 9 & 0  & 0\\
    M & 2 & 14 & 3\\
    G & 0 & 0  & 1\\
    \hline
    \end{tabular}
    \tablecomments{EE87 is~\cite{elmegreen1987}. The five galaxies with different spiral-arm classes in this work and~\cite{elmegreen1987} are  shown in Figure~\ref{fig_differ_class}.}
\end{table*}


\clearpage


\bibliography{sample63}{}
\bibliographystyle{aasjournal}

\end{document}